\documentclass{ws-ijmpe}
\usepackage[super,compress]{cite}
\usepackage{color}
\begin{document}

\markboth{Miao, Xia, Lai, Maruyama, Xu, and Zhou}{A bag model of condensed matter by the strong interaction}

\catchline{}{}{}{}{}
%%%%%%%%%%%%%%%%%%%%%%%%%%%%%%%%%%%%%%%%%%%%%%%%%%%%%%%%%%%%%%%%%%%%

\title{A bag model of matter condensed by the strong interaction}

\author{Zhi-Qiang Miao}
\address{Department of Astronomy, Xiamen University, Xiamen, Fujian 361005, China}

\author{Cheng-Jun Xia}
\address{Center for Gravitation and Cosmology, College of Physical Science and Technology, Yangzhou University, Yangzhou 225009, China\\
School of Information Science and Engineering, NingboTech University, Ningbo 315100, China\\
Advanced Science Research Center, Japan Atomic Energy Agency, Shirakata 2-4, Tokai, Ibaraki 319-1195, Japan\\
{cjxia@yzu.edu.cn}}

\author{Xiao-Yu Lai}
\address{Department of Physics and Astronomy, Hubei University of Education, Wuhan 430205, China}

\author{Toshiki Maruyama}
\address{Advanced Science Research Center, Japan Atomic Energy Agency, Shirakata 2-4, Tokai, Ibaraki 319-1195, Japan}

\author{Ren-Xin Xu}
\address{School of Physics and State Key Laboratory of Nuclear Physics and Technology, Peking University, Beijing 100871, China\\
Kavli Institute for Astronomy and Astrophysics, Peking University, Beijing 100871, China\\
r.x.xu@pku.edu.cn}

\author{En-Ping Zhou}
\address{School of Physics, Huazhong University of Science and Technology, 1037 Luoyu Road, Wuhan, 430074, China}

\maketitle

\begin{history}
\received{18 11 2021}
%\revised{Day Month Year}
%\accepted{Day Month Year}
%\comby{(xxxxxxxxxx)}
\end{history}

\begin{abstract}
Inspired by various astrophysical phenomenons, it is suggested that pulsar-like compact stars are comprised entirely of strangeons (quark-clusters with three-light-flavor symmetry) and a small amount of electrons. In order to better constrain the properties of strangeon stars, we propose a linked bag model to describe the condensed matter by the strong interaction (i.e., strong condensed matter) in both 2-flavoured (nucleons) and 3-flavoured (hyperons, strangeons, etc.) scenarios. The model parameters are calibrated to reproduce the saturation properties of nuclear matter, which are later applied to hyperon matter and strangeon matter. Compared with baryon matter, the derived energy per baryon of strangeon matter is reduced if the strangeon carries a large number of valence quarks, which stiffens the equation of state and consequently increases the maximum mass of strangeon stars. In a large parameter space, the maximum mass and tidal deformability of strangeon stars predicted by the linked bag model are consistent with the current astrophysical constraints. It is found that the maximum mass of strangeon stars can be as large as $\sim 2.5M_\odot$, while the tidal deformability of a $1.4M_\odot$ strangeon star lies in the range of $180\lesssim \Lambda_{1.4} \lesssim 340$.
\end{abstract}

\keywords{bag model; compact stars.}

\ccode{PACS numbers: 21.30.Fe, 26.60.+c, 21.65.+f}

%21.30.Fe Forces in hadronic systems and effective interactions
%21.65.+f Nuclear matter
%26.60.+c Nuclear matter aspects of neutron stars

%\tableofcontents

\section{Introduction}

What is the state of matter if normal baryon matter is compressed so tightly that baryons come into close contact? This question is not only relevant to low-energy strong force, as in the case of nuclear physics, but also important to unveil an interesting piece of Nature: the huge and dense lump created in a core-collapse supernova. The density of matter in the lump is extremely high, which may even surpass $5n_0$ in the center region with $n_0$ being the nuclear saturation density.
Two questions are frequently raised in the study of such core-compressed matter~\cite{Li2020}: 1. Does deconfinement phase transition take place~\cite{Weissenborn2011_ApJ740-L14, Klahn2013_PRD88-085001, Zhao2015_PRD92-054012, Kojo2015_PRD91-045003, Li2015_PRC91-035803, Masuda2016_EPJA52-65, Whittenbury2016_PRC93-035807, Bastian2018_Universe4-67, Annala2020_NP, Blaschke2020_Universe6-81, Blaschke2020_EPJA56-124, Xia2020_PRD102-023031}? 2. Does strangeness play an important role~\cite{Weissenborn2012_PRC85-065802, Bednarek2012_AA543-A157, Oertel2015_JPG42-075202, Maslov2016_NPA950-64, Takatsuka_EPJA13-213, Lonardoni2015_PRL114-092301, Togashi2016_PRC93-035808, Vidana2015_AIPCP1645-79, Fortin2017_PRC95-065803, Sun2018_CPC42-25101, Holdom2018_PRL120-222001, Sun2019_PRD99-023004, Zhao2019_PRD100-043018, Zhang2020_PRD101-043003}?

Since it is still challenging to simulate dense matter with lattice QCD, to answer those questions, we need to rely on various constraints from both nuclear and astrophysical studies. So far, the properties of nuclear matter around the saturation density ($n_0\approx 0.16\ \mathrm{fm}^{-3}$) are well constrained with the binding energy $B\approx -16$ MeV, the incompressibility $K = 240 \pm 20$ MeV~\cite{Shlomo2006_EPJA30-23}, the symmetry energy $S = 31.7 \pm 3.2$ MeV and its slope $L = 58.7 \pm 28.1$ MeV~\cite{Li2013_PLB727-276, Oertel2017_RMP89-015007}. Combining them with the data from PREX-II~\cite{PREX2021_PRL126-172502}, chiral effective field theory, and heavy ion collisions, more stringent constraints can be obtained~\cite{Zhang2020_PRC101-034303, Essick2021_PRL127-192701}. The observational masses and radii of PSR J0030+0451 and PSR J0740+6620~\cite{Riley2019_ApJ887-L21, Riley2021_ApJ918-L27, Miller2019_ApJ887-L24, Miller2021_ApJ918-L28, Fonseca2021_ApJ915-L12}, as well as the tidal deformability from the neutron star merger event {GRB} 170817A-{GW}170817-{AT} 2017gfo~\cite{LVC2018_PRL121-161101} will shed light on the properties of stellar matter at larger densities.

For gravity-bound stars, by combining astrophysical observations and theoretical ab initio calculations in a model-independent way, it was shown that the inferred properties of matter in the interior of most massive compact stars exhibits characteristics of the deconfined phase~\cite{Annala2020_NP}. Nevertheless, a strong first-order phase transition may be excluded due to the similarity in the radii of PSR J0030+0451 and PSR J0740+6620 despite their large differences in mass~\cite{Pang2021_ApJ922-14}. The role of strangeness in compact stars were also examined extensively~\cite{Weissenborn2012_PRC85-065802, Bednarek2012_AA543-A157, Oertel2015_JPG42-075202, Maslov2015_PLB748-369, Maslov2016_NPA950-64, Takatsuka_EPJA13-213, Vidana2011_EPL94-11002, Yamamoto2013_PRC88-022801, Lonardoni2015_PRL114-092301, Togashi2016_PRC93-035808, Weissenborn2011_ApJ740-L14, Klahn2013_PRD88-085001, Zhao2015_PRD92-054012, Kojo2015_PRD91-045003, Masuda2016_EPJA52-65, Li2015_PRC91-035803, Whittenbury2016_PRC93-035807, Fukushima2016_ApJ817-180, Sun2018_CPC42-25101, Sun2019_PRD99-023004, Dexheimer2021_PRC103-025808, Tu2022_ApJ925-16}, where additional repulsive interaction needs to be introduced to avoid the Hyperon Puzzle~\cite{Vidana2015_AIPCP1645-79}.

Normal atomic nucleus is 2-flavoured ($u$ and $d$), but ``giant nucleus'' at supra-nuclear densities may very well lie in the regime of 3 flavours of quarks ($u$, $d$, and $s$). It is thus proposed that the core-collapse compressed matter could actually be strange matter, either strange quark matter (quarks free, e.g., Refs.~\cite{Witten1984_PRD30-272, Haensel1986_AA160-121, Alcock1986_ApJ310-261, Weber2005_PPNP54-193}) or strangeon matter (quarks localized almost in a certain unit, called strangeon~\cite{Xu2003_ApJ596-L59, Wang2017_ApJ837-81}). The compact stars comprised of those matter are bound by strong force, which leads to a sharp decrease of density and results in a bare surface, i.e., strong-bound stars.

In principle, a strangeon is a colorsinglet $N_{\rm q}$-quark state with the number of quarks $N_{\rm q}=6$, 9, 12, 15, and 18, which includes same amounts of $u$, $d$, and $s$ quarks. Due to the non-observation of those multi-quark states, a strangeon may not be stable or only weakly bound in vacuum according to various investigations~\cite{Jaffe1977_PRL38-195, Jaffe1977_PRL38-617, Aerts1978_PRD17-260, Maltm1992_PLB291-371, Maezawa2005_PTP114-317, Lee2009_EPJC64-283, Beane2011_PRL106-162001, Inoue2011_PRL106-162002, Sasaki2020_NPA998-121737}. However, if strangeons are compressed tightly together, the corresponding strangeon matter may become stable due to the strong attractive interactions~\cite{Sakai1997_NPA625-192, Glendenning1998_PRC58-1298, Lai2013_MNRAS431-3282}, which could form compact stars called strangeon stars~\cite{Xu2018_SCPMA61-109531}.
%Wetzorke2003: I. Wetzorke, F. Karsch, Nucl. Phys. Proc. Suppl. 119 (2003) 278.
Astrophysically, observational consequences of strangeon stars show that various manifestations of pulsar-like compact objects could be interpreted in the regime of strangeon stars~\cite{Horvath2005_MPLA20-2799, Owen2005_PRL95-211101, Mannarelli2007_PRD76-074026, Lai2017_JPCS861-012027, Lai2019_EPJA55-60, Lai2021_RAA21-250, Wang2020_MNRAS500-5336, Gao2021_MNRAS}, to be tested by future advanced facilities (e.g., FAST, SKA, and eXTP).
Both nuclear matter (2-flavoured) and strangeon matter (3-flavoured) can be regarded as strong condensed-matter, which is simply termed as strong matter~\cite{Xu2018_SCPMA61-109531}.

The properties of strangeons in vacuum (i.e., H-dibaryons, strange tribaryons, etc.) were investigated extensively based on various methods. For example, their masses were obtained with QCD-inspired models, i.e., the MIT bag model~\cite{Jaffe1977_PRL38-195, Jaffe1977_PRL38-617, Aerts1978_PRD17-260, Mulders1980_PRD21-2653, Liu1982_PLB113-1, Maltm1992_PLB291-371, Maezawa2005_PTP114-317}, nonrelativistic quark cluster model~\cite{Oka1983_PLB130-365, Straub1988_PLB200-241, Oka1991_NPA524-649, Shen1999_JPG25-1807}, Skyrme model~\cite{Balachandran1984_PRL52-887, Jaffe1985_NPB258-468, Yost1985_PRD32-816, Kopeliovich1992_NPA549-485}, diquark model~\cite{Lee2009_EPJC64-283}, and so on. In recent years, the properties of H-dibaryons were investigated with lattice QCD close to the physical $\pi$ mass~\cite{Beane2011_PRL106-162001, Inoue2011_PRL106-162002, Sasaki2020_NPA998-121737}.
However, as the number of quarks in a strangeon $N_\mathrm{q}$ increases, the numerical cost of lattice QCD grows drastically. Similar situation is expected for nonrelativistic quark cluster model since the number of basises grows exponentially.

The properties of strangeons in dense medium, on the other hand, are even less known. In our previous investigations, strangeons inside strangeon stars were treated as individual particles, while their interactions were taken from phenomenological potential models~\cite{Lai2009_MMRAS398-L31, Lai2013_MNRAS431-3282}. In this work, we attempt to obtain both the properties of strangeons and their interactions in a unified manner.

The interaction between nucleons inside nuclei was studied assuming nucleons to be bag-like.
For infinite strong matter with negligible surface effect, the interactions between two or more bags can be accounted for if the bags are connected, i.e., a linked bag model, or a bag crystal model~\cite{Zhang1992_PRC46-2294}. The dynamics of quark propagation between separated bags would thus introduce effective interactions so that ``bags'' are condensed in strong-matter. In such cases, we adopt the MIT bag model~\cite{Chodos1974_PRD9-3471} to investigate the 2- and 3-flavoured strong matter in a unified manner, where quarks are assumed to be free in a bag-like hadron (perturbative QCD vacuum inside) immersed in a QCD vacuum characterized by a bag constant $B$. By carefully calibrate the model parameters, as will be illustrated in this work, the properties of nuclear matter, hyperon matter, and strangeon matter can be obtained simultaneously.

This paper is organized as follows. In Sec.~\ref{sec:model}, we introduce the basic framework of the linked bag model. The model is then applied to investigate the properties of nuclear matter, hyperon matter, and strangeon matter in Sec.~\ref{sec:results}, where the model parameters are fixed according to the saturation properties of nuclear matter. With the obtained equation of states (EOSs), the structures of neutron stars, hyperon stars, and strangeon stars are examined and confronted with astrophysical observations. We draw our conclusions in Sec.~\ref{sec:conclusion}.

%-m-deloi-n tan- -- bag --|---------|---------|---------|---------|---------|---------|---------|
\section{The linked bag model}\label{sec:model}
%___________________________
%\subsection{The linked bag model}\label{subsec:model}

\begin{figure}
%\hspace{-3cm}
\centering
\includegraphics[width=0.6\linewidth]{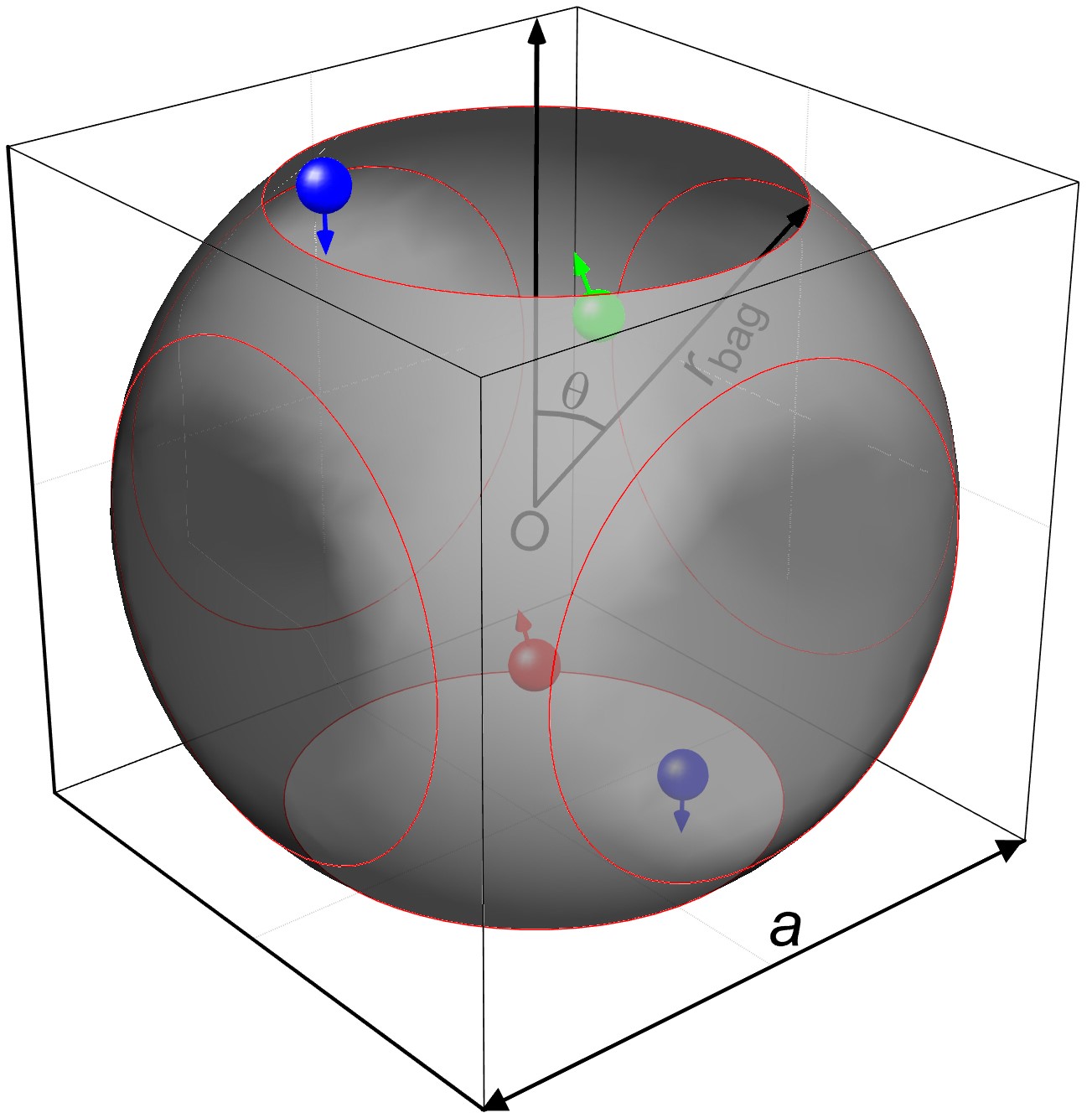}
\caption{A schematic illustration of a lattice cell in strong matter, in which point ``O'' is the center of the cell. A spherical bag (centered at ``O'' too) is inside the cell, which is linked to other ones through the six windows (the red circles plotted) on the bag's surface. The size of each window is characterized by the angle $\theta$ with $\cos\theta = a/(2r_{\rm bag})$, where $a$ is the lattice constant and $r_{\rm bag}$ the bag radius.}
\label{fig:bag lattice}
\end{figure}

In the linked bag model scenario, strong matter is comprised of quark bags with radius $r_{\rm bag}$ and quark number $N_{\rm q}$. For simplicity, we assume that the bags arrange themselves in simple cubic lattices. The lattice constant $a$ is related to the baryon number density $n$ by $a=(A/n)^{1/3}$, where $A=N_{\rm q}/3$ is the baryon number of a single bag in a lattice cell.
If $r_{\rm bag}>a/2$, the bags overlap with each other, and those six parts beyond the cell in Fig.~\ref{fig:bag lattice} are cut off since they are connected with adjacent cells, leaving behind the main part of the bag with six windows on the surface. The open angle of the window is defined as $\theta=\arccos(a/2r_{\rm bag})$. Obviously, the bag surface will disappear when $r_{\rm bag}\geq \sqrt{3}a/2$ (i.e., $\theta\geq54.7^\circ$), implying that strong matter may undergo a deconfinement phase transition.

Instead of solving the Dirac equations for quarks, we adopt the Fermi-gas approximation while keeping other terms from the MIT bag model. The energy per lattice cell is obtained with
\begin{equation}
\label{eq:energy per cell}
E=\sum_{j}(\Omega_j+N_j\mu_j)+BV-\frac{z_0}{r_{\rm bag}}\frac{\omega}{4\pi},
\end{equation}
where $\Omega_j$, $N_j$ and $\mu_j$ denote the thermodynamic potential, total particle number, and chemical potential of particle type $j$. Here after we use $j$ for both quarks and electrons, while $i$ only for quarks, $B$ for the bag parameter, and $V$ for the enclosed volume of the bag. The third term of Eq.~(\ref{eq:energy per cell}) corresponds to the zero-point energy in traditional MIT bag model~\cite{DeGrand1975_PRD12-2060}, while here we readjust $z_0$ to compensate the energy shift in Fermi-gas approximations as well.
The variable $\omega$ represents the solid angle of the remaining bag, which is obtained at given $r_{\rm bag}$ and $a$, i.e.,
\begin{equation}
  \omega =
 \left\{\begin{array}{l}
   4\pi,\  r_{\rm bag}<\frac{a}{2}\\
   4\pi\left(\frac{3a}{2r_{\rm bag}}-2\right),\  \frac{a}{2}\leq r_{\rm bag}<\frac{\sqrt{2}a}{2}\\
   \int_{\theta_1}^{\theta_2} \cos \theta \left[\frac{\pi}{2}-2\cos^{-1}\left(\frac{a}{2r_{\rm bag}\cos \theta}\right)\right]\mbox{d}\theta,\  \frac{\sqrt{2}a}{2}\leq r_{\rm bag}<\frac{\sqrt{3}a}{2}\\
   0,\  r_{\rm bag}\geq\frac{\sqrt{3}a}{2}\\
 \end{array}\right., \label{Eq:omega}
\end{equation}
where $\theta_1=\cos^{-1}\left(\frac{a}{\sqrt{2}r_{\rm bag}}\right)$ and $\theta_2=\sin^{-1}\left(\frac{a}{{2}r_{\rm bag}}\right)$.
In the extreme case of isolated bags, we have $\omega=4\pi$ and the dimensionless parameter $z_0$ is fixed by fitting to hadron spectra. The solid angle $\omega$ starts to decrease from $4\pi$ when the bags are linked as indicated in Fig.~\ref{fig:bag lattice}. Once $r_{\rm bag}$ reaches $\sqrt{3}a/2$, $\omega$ vanishes and the bag takes up the entire volume of the lattice cell with $V=a^3$, i.e., a deconfinement phase transition that restores Eq.~(\ref{eq:energy per cell}) into its original MIT bag model description of quark matter. Since we have adopted the Fermi-gas approximation instead of solving the quark single particle energies exactly, the parameter $z_0$ needs to vary with density to restore the discrete levels, i.e., $z_0=z_0(n)$. In practice, we fix $z_0(n)$ by reproducing the saturation properties of nuclear matter.

The finite-size effects of the linked bag is treated with the multiple reflection expansion (MRE) method~\cite{Berger1987_PRC35-213, Berger1991_PRC44-566, Madsen1994_PRD50-3328}, where the thermodynamic potential in Eq.~(\ref{eq:energy per cell}) is expanded as
\begin{equation}
\Omega_i=\Omega_{i,V}V+\Omega_{i,S}S+\Omega_{i,C}C.
\end{equation}
Here the volume ($\Omega_{i,V}$), surface ($\Omega_{i,S}$), and curvature ($\Omega_{i,C}$) contributions are given by~\cite{Fraga2005_PRD71-105014, Berger1987_PRC35-213, Berger1991_PRC44-566, Madsen1994_PRD50-3328}
\begin{eqnarray}
\Omega_{i,V}&=& -\frac{g_i}{24\pi^2}\left[\mu_i u_i(\mu_i^2-\frac52m_i^2)+\frac32m_i^4\ln\frac{\mu_i+u_i}{m_i}\right]\label{eq:Omega_V}\\
            &&{}+\frac{g_i\alpha_s}{12\pi^3}\left[3\left(\mu_i m_i-m_i^2\ln\frac{\mu_i+u_i}{m_i}\right)^2-2u_i^4\right.\nonumber\\
            &&{}\left.+\left(6m_i^2\ln\frac{\bar\Lambda}{m_i}+4m_i^2\right)\left(\mu_i u_i - m_i^2 \ln\frac{\mu_i+u_i}{m_i}\right)\right], \nonumber \\
\Omega_{i,S}&=&\frac{g_i}{8\pi}\left[\frac{\mu_i u_i^2}{6}-\frac{1}{3\pi}\left(\mu_i^3\arctan\frac{u_i}{m_i}-2\mu_i u_i m_i  + m_i^3\ln\frac{\mu_i+u_i}{m_i}\right)\right.\nonumber\\
            &&{}\left.-\frac{m_i^2(\mu_i-m_i)}{3} \right], \label{eq:Omega_S} \\
\Omega_{i,C}&=&\frac{g_i}{48\pi^2}\left(m_i^2\ln\frac{\mu_i+u_i}{m_i}+\frac{\pi}{2}\frac{\mu_i^3}{m_i}-\frac{3\pi\mu_im_i}{2}+\pi m_i^2-\frac{\mu_i^3}{m_i}\arctan\frac{u_i}{m_i}\right), \label{eq:Omega_C}
\end{eqnarray}
with $u_i\equiv\sqrt{\mu_i^2-m_i^2}$ and $g_i$ the degeneracy factor ($g_u=g_d=g_s=6$) for quark flavor $i$. The area $S$ and curvature $C$ of the bag are obtained with $S=\omega r_{\rm bag}^2$ and $C=2\omega r_{\rm bag}$, respectively. Note that in Eq.~(\ref{eq:Omega_V}) we have considered the first-order correction to the thermodynamic potential
of QCD. The coupling constant $\alpha_s$ and quark masses $m_i$ are running with energy scale~\cite{Fraga2005_PRD71-105014}, i.e.,
\begin{eqnarray}
\alpha_s(\bar\Lambda) &=&\frac{1}{\beta_0\mathfrak{L}}\left(1-\frac{\beta_1\ln \mathfrak{L}}{\beta_0^2\mathfrak{L}}\right), \\
m_i(\bar\Lambda)&=&\hat m_i\alpha_s^{\gamma_0/\beta_0}\left[1+\left(\frac{\gamma_1}{\beta_0}-\frac{\beta_1\gamma_0}{\beta_0^2}\right)\alpha_s\right],
\end{eqnarray}
where $\mathfrak{L}=2\ln(\bar\Lambda/\Lambda_{\overline {\rm MS}})$ and $\Lambda_{\overline {\rm MS}}$ is the ${\rm \overline{MS}}$ renormalization point. In this work we take $\Lambda_{\overline {\rm MS}}=376.9\,{\rm MeV}$ and $\hat m_u=\hat m_d=0$, $\hat m_s=220$ and 280 MeV. The parameters of $\beta$-function and $\gamma$-function are $\beta_0=\frac{1}{4\pi}(11-\frac23N_{\rm f})$, $\beta_1=\frac{1}{16\pi^2}(102-\frac{38}{3}N_{\rm f})$, $\gamma_0=1/\pi$ and $\gamma_1=\frac{1}{16\pi^2}(\frac{202}{3}-\frac{20}{9}N_{\rm f})$ with $N_{\rm f}=3$~\cite{Vermaseren1997_PLB405-327}. The renormalization scale envolves with the chemical potentials of quarks, and we adopt $\bar\Lambda=\frac{C_1}{3}\sum_i\mu_i$ with $C_1=1\sim 4$~\cite{Fraga2014_ApJ781-L25}.

The bag parameter $B$ was introduced to account for the energy difference between the physical and perturbative vacua~\cite{Chodos1974_PRD9-3471}. According to QCD sum-rule~\cite{Shuryak1978_PLB79-135}, one finds $B\simeq455\,{\rm MeV/fm^3}$ at vanishing chemical potentials, while fitting to the hadron spectra gives a lower value $B\simeq 50\,{\rm MeV/fm^3}$~\cite{DeGrand1975_PRD12-2060}. At larger chemical potentials, however, it is found that $B$ prefers a larger value by comparing with the pQCD calculations to higher orders~\cite{Fraga2014_ApJ781-L25}. To account for these values in our current study, we assume $B$ varies with chemical potential and take a third-order expansion with respect to $\xi$, i.e.,
\begin{equation}
B=B_0+B_2\xi^2+B_3\xi^3,
\end{equation}
where $\xi=(\sum_i N_i\mu_i/A-m_N)/m_N$ with $A$ ($=\sum_i N_i/3$) being the baryon number of each lattice cell and $m_N = 938$ MeV the nucleon mass. This expansion is composed by three parts: the constant part $B_0$, the symmetric part $B_2\xi^2$ and the asymmetric part $B_3\xi^3$. Note that the first-order term is discarded so that ${\partial B}/{\partial \mu_i}=0$ at $\xi=0$. The nuclear symmetry energy are then accounted for by taking such a form for $B$, since the contribution from perturbative interaction and kinetic energy of quarks does reach the experimental value of symmetry energy if we take $B_2=B_3=0$. In this work we fix $B=B_0=50\,{\rm MeV/fm^3}$ at $\sum_i N_i\mu_i/A=m_N$ ($\xi=0$). The remaining parameters $B_2$ and $B_3$ are left undetermined and will be fixed later. The particle number $N_j$ is then related to the chemical potentials $\mu_j$ via
\begin{equation}
\label{eq:Ni}
N_j=-\frac{\partial \Omega_j}{\partial \mu_j}-\frac{\partial B}{\partial \mu_j}V.
\end{equation}

The bag radius $r_{\rm bag}$ is then fixed by minimizing the total energy $E$ at given cell volume $a^3$ and particle numbers $N_i$. With the energy per baryon determined by $E/A$, the energy density reads
\begin{equation}
\label{eq:energy density}
\varepsilon=nE/A.
\end{equation}
According to the basic thermodynamic relations, the baryon chemical potential and pressure are obtained with
\begin{eqnarray}
\mu_{\rm b}&=&\frac{d \varepsilon}{d n}, \\
P&=&n^2\frac{d}{d n}\frac{\varepsilon}{n}=n \mu_{\rm b}-\varepsilon.
\label{eq:eos}
\end{eqnarray}

%------ s--|---------|---------|---------|---------|---------|---------|---------|
\section{Strong matter and compact stars}\label{sec:results}

\subsection{Model parameters}\label{subsec:param}
In our current study, the energy per baryon of both symmetric nuclear matter and neutron matter is obtained by taking $N_u=N_d=3/2$ and $N_u=N_d/2=1$, respectively. At given $B_2$ and $B_3$, the model parameters $C_1$ and $z_0(n_0)$ are fixed by reproducing the saturation properties of nuclear matter, while $z_0(n)$ at $n\neq n_0$ is obtained by fitting to the energy per baryon of symmetric nuclear matter. In particular, adopting the linked bag model, we reproduce the energy per baryon of nuclear matter predicted by
\begin{equation}
\label{eq:NM}
E_\mathrm{NM}=E_0(n_0)+\frac{K_0}{2}(\frac{n-n_0}{3n_0})^2+E_{\rm sym}(n)\delta^2
\end{equation}
with the symmetry energy
\begin{equation}
E_{\rm sym}(n)=E_{\rm sym}(n_0)+L(\frac{n-n_0}{3n_0}).
\end{equation}
Here $\delta=(n_n-n_p)/n=N_d-N_u$ represents the isospin asymmetry with $n_p$ and $n_n$ being the proton and neutron number densities. According to various experimental investigations and nuclear theories~\cite{Shlomo2006_EPJA30-23, Li2013_PLB727-276, Oertel2017_RMP89-015007}, the parameters in Eq.~(\ref{eq:NM}) are well constrained. We thus take $n_0=0.16\,{\rm fm^{-3}}$, $E_0(n_0)= 922$ MeV, $K_0=240\,{\rm MeV}$ and $E_{\rm sym}(n_0)=31.7\,{\rm MeV}$, while several values of $L$ are adopted due to its larger uncertainty.

%__________________________________
\begin{figure}
\centering
\includegraphics[width=\linewidth]{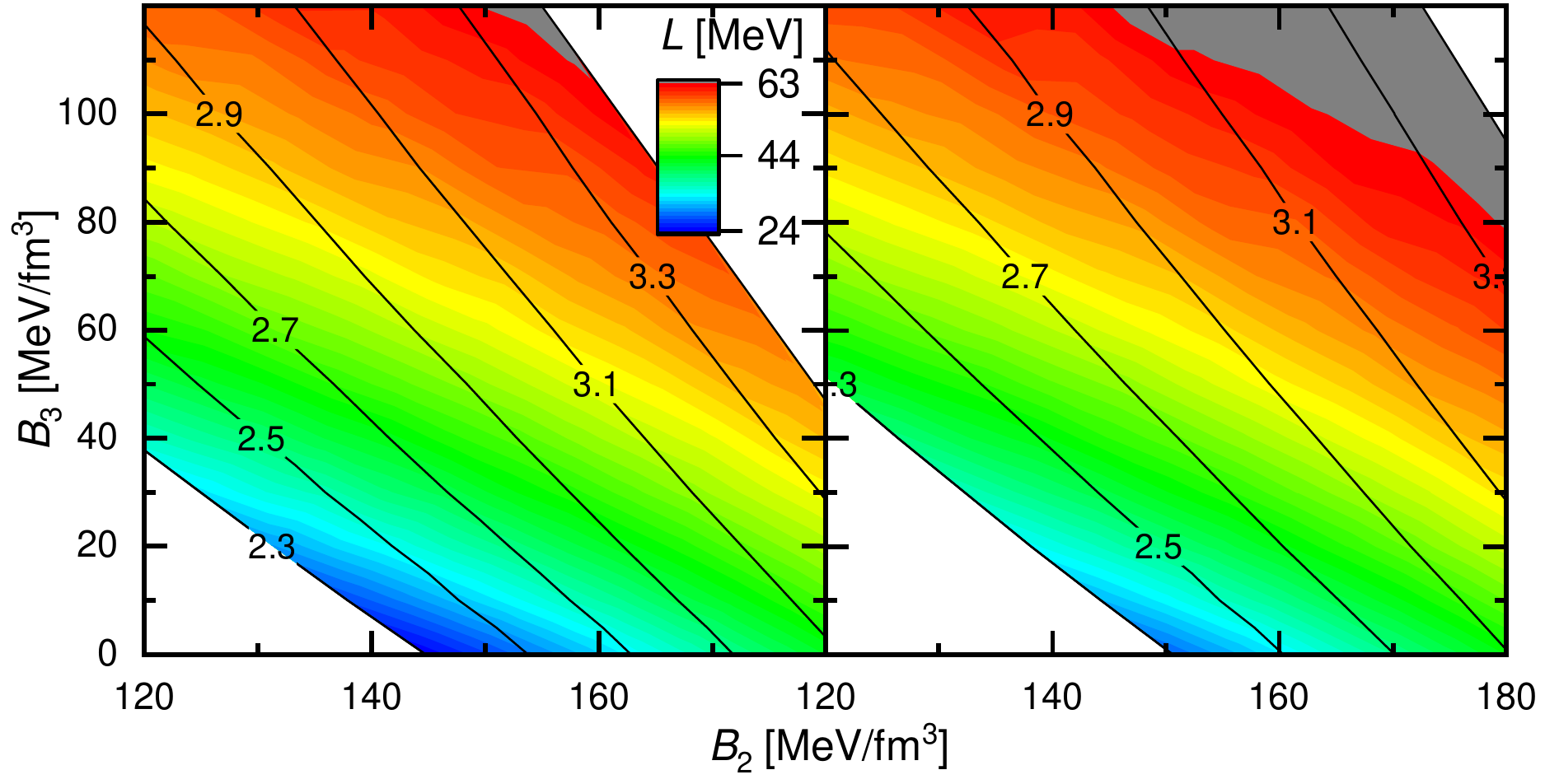}
\caption{The slope $L$ of symmetry energy and $B_3$ obtained with various combinations of parameters $B_2$, $C_1$ (indicated with solid curves), and two invariant strange quark masses $\hat m_s = 220\,{\rm MeV}$ (Left) and 280 MeV (Right). }
\label{fig:b2b3 para space}
\end{figure}
%_____________________________________________

At given $B_2$ and $B_3$, the parameter $z_0$ in the linked bag model is fixed by reproducing the energy per baryon of symmetric nuclear matter obtained with Eq.~(\ref{eq:NM}), which is essentially density dependent and connected to the incompressibility parameter $K_0=240$ MeV. Once we fix $z_0(n)$, the parameter $C_1$ is determined by reproducing the symmetry energy $E_{\rm sym}(n_0)=31.7$ MeV. Note that the slope of symmetry energy $L(n_0)$ is essentially determined by $B_2$ and $B_3$, so that the parameters $B_2$ and $B_3$ can be better constrained if $L$ can be fixed. In Fig.~\ref{fig:b2b3 para space} we present the constraints on the parameter set $(B_2,B_3)$, where we have taken either $C_1$ (black curves) or $L$ (red curves) as constant values.

\begin{table*}
\centering
\caption{\label{tab: paras of nuclear} Parameter sets $(C_1,\hat m_s,B_2,B_3,z_0(n_0))$ chosen to reproduce saturation properties of nuclear matter: the saturation density $n_0 = 0.16\,{\rm fm^{-3}}$, the minimum energy per baryon $E_0(n_0) = 922\ {\rm MeV}$, the incompressibility $K = 240\,{\rm MeV}$, the symmetry energy $E_{\rm sym}(n_0) = 31.7\,{\rm MeV}$. The obtained slope of symmetry energy $L$ (in MeV) is listed here as well.}
\begin{tabular}{ccccccc}\hline\hline
& $C_1$ & $\hat m_s$[${\rm MeV}$] & $B_2$[${\rm MeV/fm^3}$]& $B_3$[${\rm MeV/fm^3}$] & $z_0(n_0)$ & $L$[${\rm MeV}$] \\ \hline
({\romannumeral1}) & 2.7 & 220 & 136.7 & 50 & 2.944 & 45.1 \\
({\romannumeral2}) & 2.7 & 220 & 112.7 & 100 & 2.926 & 52.7 \\
({\romannumeral3}) & 2.7 & 280 & 125.0 & 100 & 2.908 & 56.6 \\
({\romannumeral4}) & 3.2 & 280 & 162.3 & 100 & 2.843 & 62.8 \\ \hline\hline
\end{tabular}
\end{table*}

\begin{figure}
\centering
\includegraphics[width=0.6\linewidth]{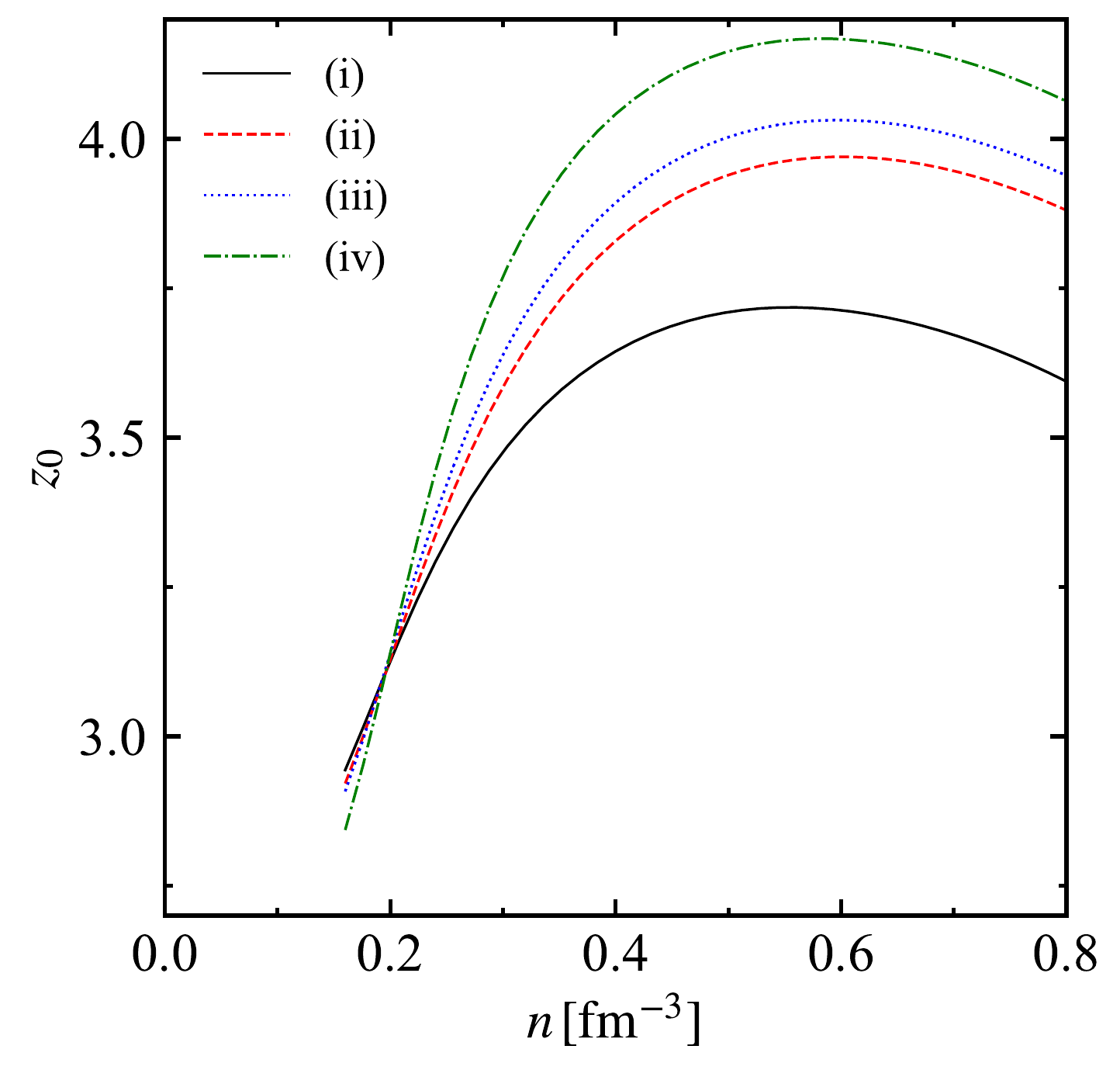}
\caption{The zero-point parameter as a function of baryon number density for the selected parameter sets listed in Table~\ref{tab: paras of nuclear}.}
\label{fig:z0 wiht n}
\end{figure}

Based on Fig.~\ref{fig:b2b3 para space}, in this work we adopt four parameter sets (i-iv) with the corresponding values listed in Table~\ref{tab: paras of nuclear}.
The obtained values of $z_0$ are presented in Fig.~\ref{fig:z0 wiht n}. It is interesting to notice that $z_0$ increases with density and reaches its peak value at $n\approx 3.5 n_0$, which later decreases at larger densities. This may be related to the variations of nucleon structures as well as the strong correlations with neighboring nucleons in nuclear medium, e.g, the EMC effect~\cite{Aubert1983_PLB123-275}. Meanwhile, the obtained nucleon radius is decreasing with density, and approaches to a constant value at highest densities.

Since one would expect that strong interactions do not vary with quark flavor, we adopt the same values of $B_0$, $B_2$, and $B_3$ as indicated in Table~\ref{tab: paras of nuclear}
for hyperon and strangeon matter with $N_{\rm q}\geq 3$ and $N_s\neq 0$. For the parameter $z_0$, keeping {it} unchanged might be reasonable for $N_{\rm q}=3$. However, $z_0(n)$ could be different at larger $N_{\rm q}$, since the quark energy level corrections should vary with respect to the quark number $N_q$. Note the total energy level correction takes an approximate form of $(k_1N_q^{4/3}-k_2N_q)/r_{\rm bag}$, where $k_1$, $k_2$ are two constants. To catch this feature of corrections, for simplicity, we rescale $z_0$ by an effective formula $z_0=(\frac{N_{\rm q}}{3})^{4/3}\tilde z_0-f$, where $\tilde z_0$ is obtained by reproducing nuclear matter properties and $f$ is a dampening factor.
We thus take $f=0$ at $N_{\rm q}=3$, i.e., $z_0 = \tilde z_0$,  while larger $f$ is expected at larger $N_{\rm q}$. The parameter $f$ is then fixed by requiring the bags to be connected, which indicates $5.8\lesssim f \lesssim 6.1$. In the discussion below, we take $f=5.8$ since $f$ would not significantly affect the properties of strangeon stars, as will be shown later in Sec.~\ref{subsec:struc of strangeon stars}.

In conclusion, comparing with traditional MIT bag model~\cite{DeGrand1975_PRD12-2060, Jaffe1977_PRL38-195, Jaffe1977_PRL38-617, Aerts1978_PRD17-260, Mulders1980_PRD21-2653, Liu1982_PLB113-1, Maltm1992_PLB291-371, Maezawa2005_PTP114-317}, we have introduced the damping parameter $f$ and the density dependent bag constant with two additional parameters $B_2$ and $B_3$ to account for the in-medium properties of strong matter. The possible combinations of those parameters are thoroughly examined in our current study, while the other parameters are taken as their typical values fitted to hadron spectra~\cite{DeGrand1975_PRD12-2060}. Note that the spin dependent interactions (e.g., the color-magnetic part of the one-gluon-exchange interaction) are not included here, which could affect the mass spectra of strangeons~\cite{Aerts1978_PRD17-260} and should be considered in our future works.

\subsection{Nucleon matter, hyperon matter, and strangeon matter}\label{subsec:prop of strangeon matt}

In this section we study strong matter inside compact stars with linked bag model, where electrons need to be included to fulfill the charge neutrality condition
\begin{equation}
\sum_i Q_i N_i+Q_e N_e=0.
\end{equation}
Here $Q_u=2/3$, $Q_d=Q_s=-1/3$ and $Q_e=-1$ are the charge of quarks and electrons. Note that electrons are not confined within the bags, the corresponding thermodynamic potential can then be obtained with
\begin{equation}
\Omega_e=\Omega_{e,{\rm V}}a^3=-\frac{\mu_e^4}{12\pi^2}a^3.
\end{equation}
In principle, $\mu^-$ will appear in the centre region of a neutron star. However, we neglect the contribution of $\mu^-$ since it is insignificant for hyperon stars and strangeon stars.

The quarks and leptons will undergo various weak reactions, i.e.,
\begin{equation}
u+e^- \rightarrow d+\nu_e,~~d \rightarrow u+e^-+\bar\nu_e.
\end{equation}
If strangeness is involved (3-flavored matter), the following reactions take place, i.e.,
\begin{subequations}
\begin{align}
u+e^-&\rightarrow s+\nu_e,~~s \rightarrow u+e^-+\bar\nu_e,\\
s+u&\leftrightarrow d+u.
\end{align}
\end{subequations}
Then the $\beta$-equilibrium is reached, i.e.,
\begin{equation}
\mu_u+\mu_e=\mu_d=\mu_s.
\end{equation}
In this work, the $\beta$-equilibrium condition is satisfied by minimizing the total energy with respect to the particle numbers $N_i$ at a given total baryon number $A = \sum_i N_i/3=N_{\rm q}/3$. Then the energy density and pressure are obtained with Eqs.~(\ref{eq:energy density}-\ref{eq:eos}), which correspond to the EOS of strong matter. By taking $N_{\rm q}=3$, the EOSs of nuclear matter and hyperon matter can be obtained, while larger $N_{\rm q}$ indicates strangeon matter. In particular, we take  $\hat m_s\rightarrow \infty$ for nuclear matter so that $s$ quark does not emerge for $\beta$-equilibrated matter, while for hyperon matter the values indicated in Table~\ref{tab: paras of nuclear} are adopted. In this paper, we limit our discussions for strong matter with $N_{\rm q}=3$ and $N_{\rm q}=9$. For all cases, as illustrated in Sec.~\ref{subsec:param}, we keep $B$ unchanged and $z_0=(\frac{N_{\rm q}}{3})^{4/3}\tilde z_0-f$ with $f$ being the dampening factor. Particularly, we take $f=0$ for $N_{\rm q}=3$ and $f=5.8$ for $N_{\rm q}=9$.

\begin{figure*}
\includegraphics[width=0.45\linewidth]{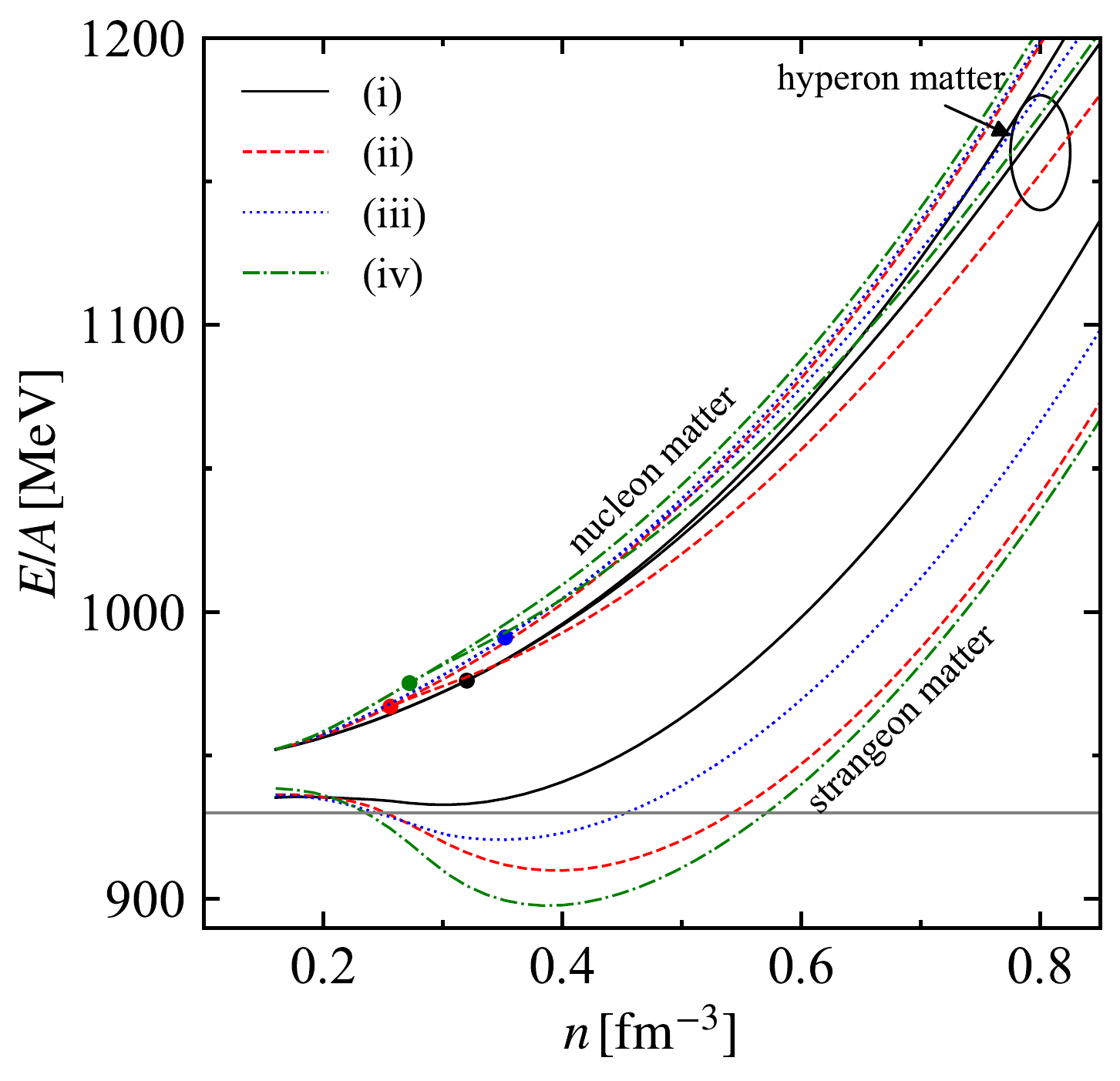}
\includegraphics[width=0.48\linewidth]{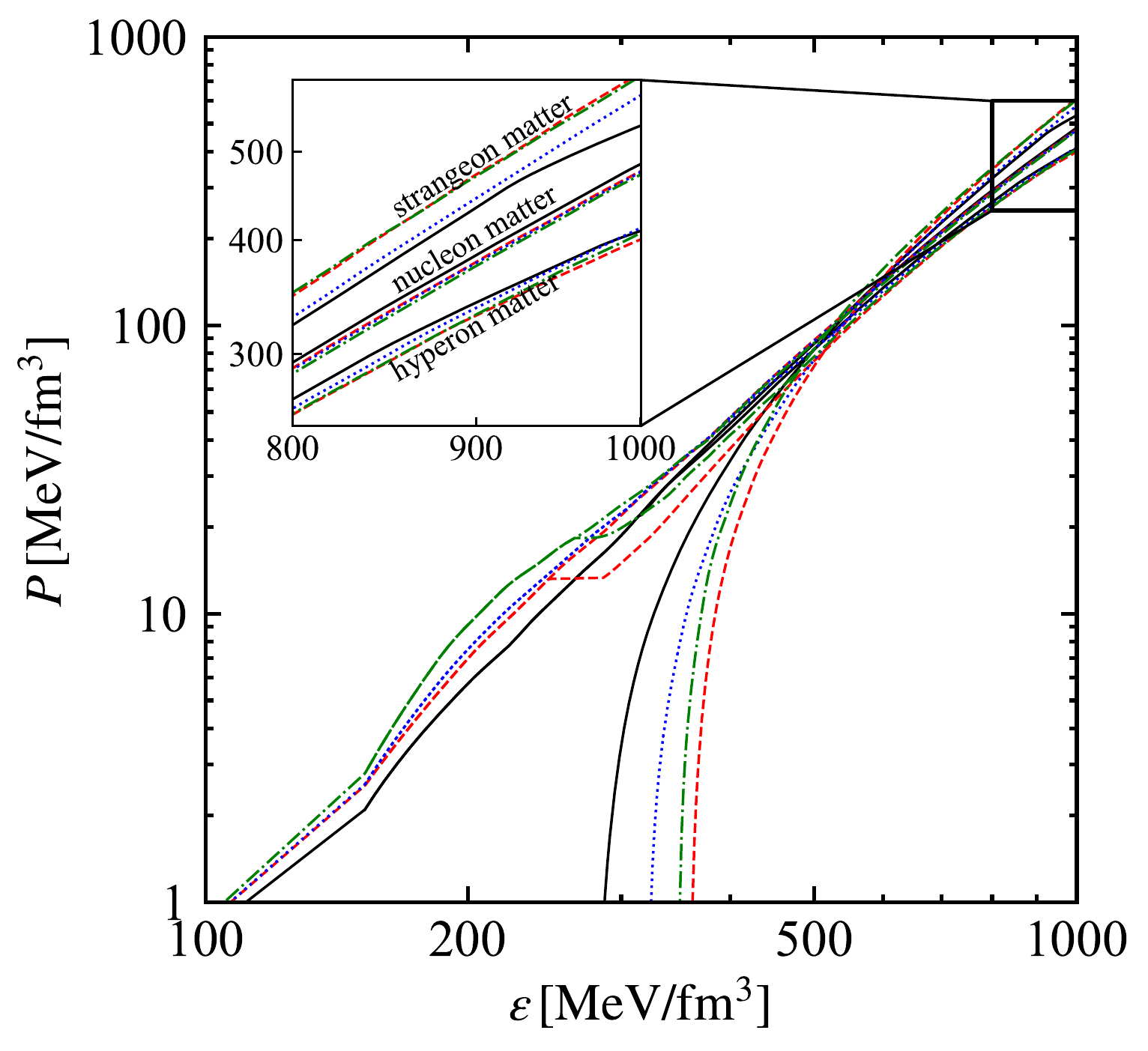}
\caption{{Energy per baryon (Left) and the corresponding EOSs (Right) for nucleon matter with $N_{\rm q}=3$, hyperon matter with $N_{\rm q}=3$, and strangeon matter with $N_{\rm q}=9$, where the parameter sets listed in Table~\ref{tab: paras of nuclear} are adopted.} For strangeon matter with $N_{\rm q}=9$, we take $f=5.8$. The horizontal line in the left panel corresponds to $E/A = 930\,{\rm MeV}$, which is the energy per baryon of the most stable atomic nucleus, $^{56}{\rm Fe}$. The solid dots indicate the critical densities at which $s$ quark starts to appear in hyperon matter.}
\label{fig:epb1 and eos1}
\end{figure*}

In Fig.~\ref{fig:epb1 and eos1} we present the energy per baryon as well as the EOSs of nuclear matter ($N_{\rm q} =3$, $N_s = 0$), hyperon matter ($N_{\rm q}=3$, $N_s\neq 0$), and strangeon matter ($N_{\rm q}=9$) in compact stars, which are obtained with the selected parameter sets in Table~\ref{tab: paras of nuclear}. In this work, our model is restricted to describe strong matter at $n\geq 0.16\,{\rm fm^{-3}}$. In the density regime of $n<0.16\,{\rm fm^{-3}}$, we employ the results of Negele \& Vautherin~\cite{Negele1973_NPA207-298} for $0.001\,{\rm fm^{-3}} < n < 0.08\,{\rm fm^{-3}}$,  and of Baym et al.~\cite{Baym1971_ApJ170-299} for $n < 0.001\,{\rm fm^{-3}}$. Between $0.08\,{\rm fm^{-3}}$ and $0.16\,{\rm fm^{-3}}$, we simply take a linear interpolation since the structures of neutron/hyperon stars are insensitive to the EOSs adopted in this density region. For each parameter set, the energy per baryon of nuclear matter is decreased once $s$-quarks (hyperons) emerge at about twice the nuclear saturation density. The energy is further reduced if we take $N_q>3$, i.e., strangeon matter with $N_q=9$. We note that strangeon matter reaches its minimum at $n=2n_0$$\sim$$3n_0$. For a few cases, the energy per baryon of strangeon matter can even be smaller than $930\,{\rm MeV}$, namely strangeon matter is more stable than $^{56}{\rm Fe}$. Combined with Fig.~\ref{fig:b2b3 para space}, it is found that the minimum energy per baryon of strangeon matter increases while the corresponding density decreases along the curves with fixed $C_1$ from top-left to lower-right regions.

In the right panel of Fig.~\ref{fig:epb1 and eos1}, it is easy to see that the EOSs of strangeon matter are stiffer than that of nuclear matter and hyperon matter, which indicates that the introduction of linked bag will results in stiffening of EOSs. The energy densities at zero pressure lie between $\sim$$280\,{\rm MeV/fm^3}$ and $\sim$$360\,{\rm MeV/fm^3}$, or, equivalently, $\sim$$1.8$ and $\sim$$2.4$ times the nuclear saturation density (mass density). It is worth noting that, although the equation of state is very stiff, the causality condition is still satisfied for strangeon matter~\cite{Lu2018_SCPMA61-089511}.

\subsection{Neutron stars, hyperon stars, and strangeon stars}\label{subsec:struc of strangeon stars}

\begin{table*}
\centering
\caption{\label{tab: strageon_star} The surface baryon number ($n_{\rm surf}$) and energy ($\varepsilon_{\rm surf}$) densities, radius ($R_{1.4}$), tidal deformability ($\Lambda_{1.4}$), TOV mass ($M_{\rm TOV}$),  and centre baryon number density ($n_{\rm c}$) for strangeon stars obtained with the parameter sets listed in Table~\ref{tab: paras of nuclear}.}
\begin{tabular}{ccccccc}\hline\hline
& $n_{\rm surf}[{\rm fm^{-3}}]$ & $\varepsilon_{\rm surf}[{\rm MeV/fm^3}]$ & $R_{1.4}[{\rm km}]$ & $\Lambda_{1.4}$ & $M_{\rm TOV}[M_\odot]$ & $n_{\rm c}[{\rm fm^{-3}}]$ \\ \hline
({\romannumeral2}) & 0.395 & 359.38 & 9.519 & 187.9 & 2.411 & 1.069 \\
({\romannumeral3}) & 0.348 & 320.56 & 9.710 & 208.8 & 2.394 & 1.086 \\
({\romannumeral4}) & 0.388 & 348.11 & 9.666 & 210.4 & 2.438 & 1.080 \\ \hline\hline
\end{tabular}
\end{table*}

The equilibrium configurations of compact stars can be obtained by solving the Tolman-Oppenheimer-Volkoff (TOV) equations for the pressure $P$ and the enclosed mass $m$, i.e.,
\begin{eqnarray}
\frac{dP}{dr}&=&-\frac{m\varepsilon}{r^2}\frac{[1+P/\varepsilon][1+4\pi r^3P/m]}{1-2m/r}, \label{eq:TOV1}\\
\frac{dm}{dr}&=& 4\pi r^2\varepsilon,
\end{eqnarray}
where $P$ and $\varepsilon$ are the pressure and energy density at the radial coordinate $r$, respectively. The dimensionless tidal deformability is related to the Love number $k_2$ through $\Lambda=\frac23k_2c^{-5}$, where $k_2$ measures how easily a star is deformed by an external tidal field and $c=M/R$ the compactness of the star~\cite{Damour2009_PRD80-084035,Hinderer2010_PRD81-123016,Postnikov2010_PRD82-024016}.

\begin{figure*}
\includegraphics[width=0.47\linewidth]{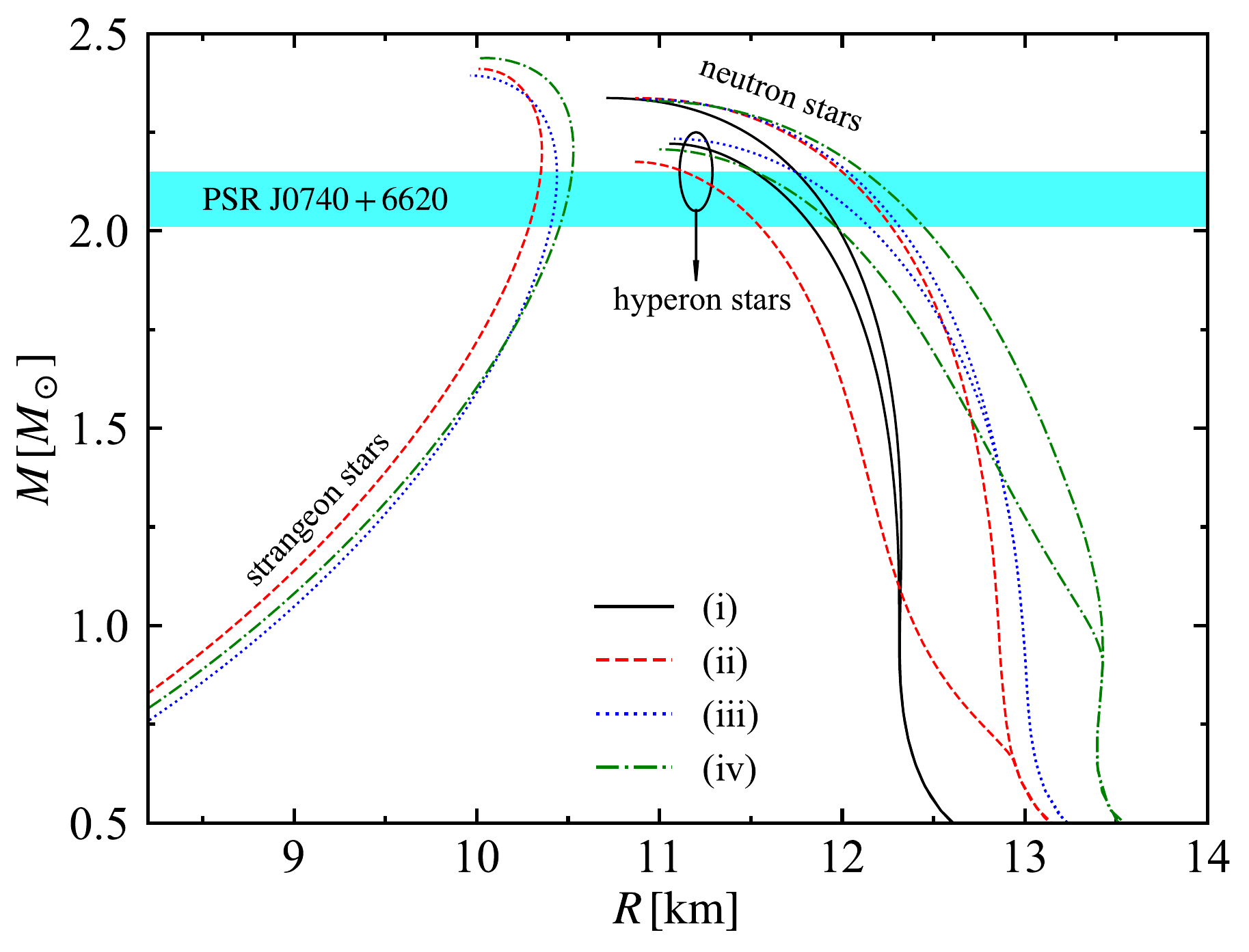}
\includegraphics[width=0.485\linewidth]{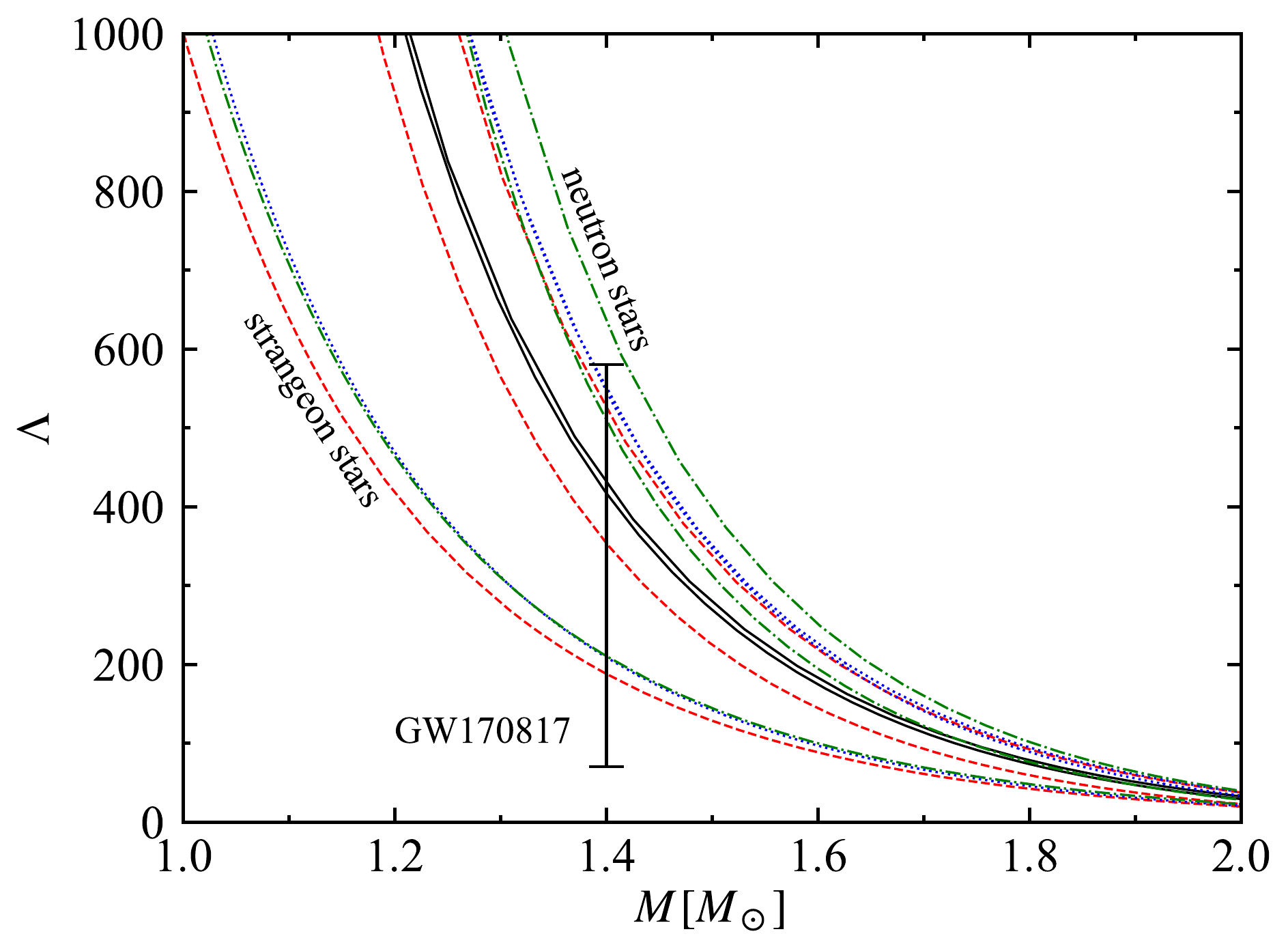}
\caption{{Mass-radius relations (Left) and tidal deformability (Right) for traditional neutron stars, hyperon stars, and strangeon stars}. For strangeon stars with $N_{\rm q}=9$, we take $f=5.8$. The observational mass of PSR J0740+6620 ($2.08 \pm 0.07\ M_\odot$)~\cite{Fonseca2021_ApJ915-L12} is indicated with the horizontal band. The LIGO/Virgo constraint~\cite{LVC2018_PRL121-161101} from GW170817 on the tidal deformability for a $1.4\,M_\odot$ star, $\Lambda_{1.4} = 190_{-120}^{+390}$, is also displayed in right panel.}
\label{fig:MR and TD1}
\end{figure*}

Based on the EOSs presented in Fig.~\ref{fig:epb1 and eos1}, the mass-radius relations of compact stars are obtained by solving Eq.~(\ref{eq:TOV1}), while the tidal deformability is determined by the second Love number $k_2$. The results are presented in Fig.~\ref{fig:MR and TD1}, where various parameter sets listed in Table~\ref{tab: paras of nuclear} are adopted. The corresponding properties of strangeon stars are listed in Table~\ref{tab: strageon_star}. In general, the maximum masses of strangeon stars are higher than those of neutron stars and hyperon stars due to the stiffer EOSs of strangeon matter. It is shown that the radius $R_{1.4}$ of a typical $1.4\,M_\odot$ star ranges from $9.5\,{\rm km}$ to $13\,{\rm km}$, where strangeon stars with $N_{\rm q}=9$ have smaller radii and larger maximum mass than those with $N_{\rm q}=3$. Note that $R_{1.4}$ generally increases with the slope of symmetry energy $L$ for neutron stars and hyperon stars~\cite{Zhu2018_ApJ862-98, Tsang2019_PLB795-533, Dexheimer2019_JPG46-034002, Zhang2019_EPJA55-39, Zhang2020_PRC101-034303, Li2020_PRC102-045807}, while such a trend is missing for strangeon stars. This is mainly due to the large surface densities ($>2n_0$) for strangeon stars without crusts, where the saturation properties of nuclear matter have little impact on their structures. It is worth mentioning that, for strangeon stars with a smaller surface density, the radii and masses generally become larger~\cite{Bhattacharyya2016_MNRAS457-3101, Li2017_ApJ844-41, Xia2021_CPC45-055104}, which is indeed the case for $R_{1.4}$ as indicated in Fig.~\ref{fig:MR and TD1}. In the right panel of Fig.~\ref{fig:MR and TD1}, we find $\Lambda$ decreases monotonously with mass{, while $\Lambda$ is reduced with the emergence of strangeness.} Except for the traditional neutron stars obtained with parameter set (iv), the tidal deformability of a typical $1.4\,M_\odot$ star $\Lambda_{1.4}$ falls in between $190$ and $550$ for all presented cases, which fulfills the GW170817 constraint of $\Lambda_{1.4}<580$~\cite{LVC2018_PRL121-161101}.

\begin{figure}
\centering
\includegraphics[width=0.7\linewidth]{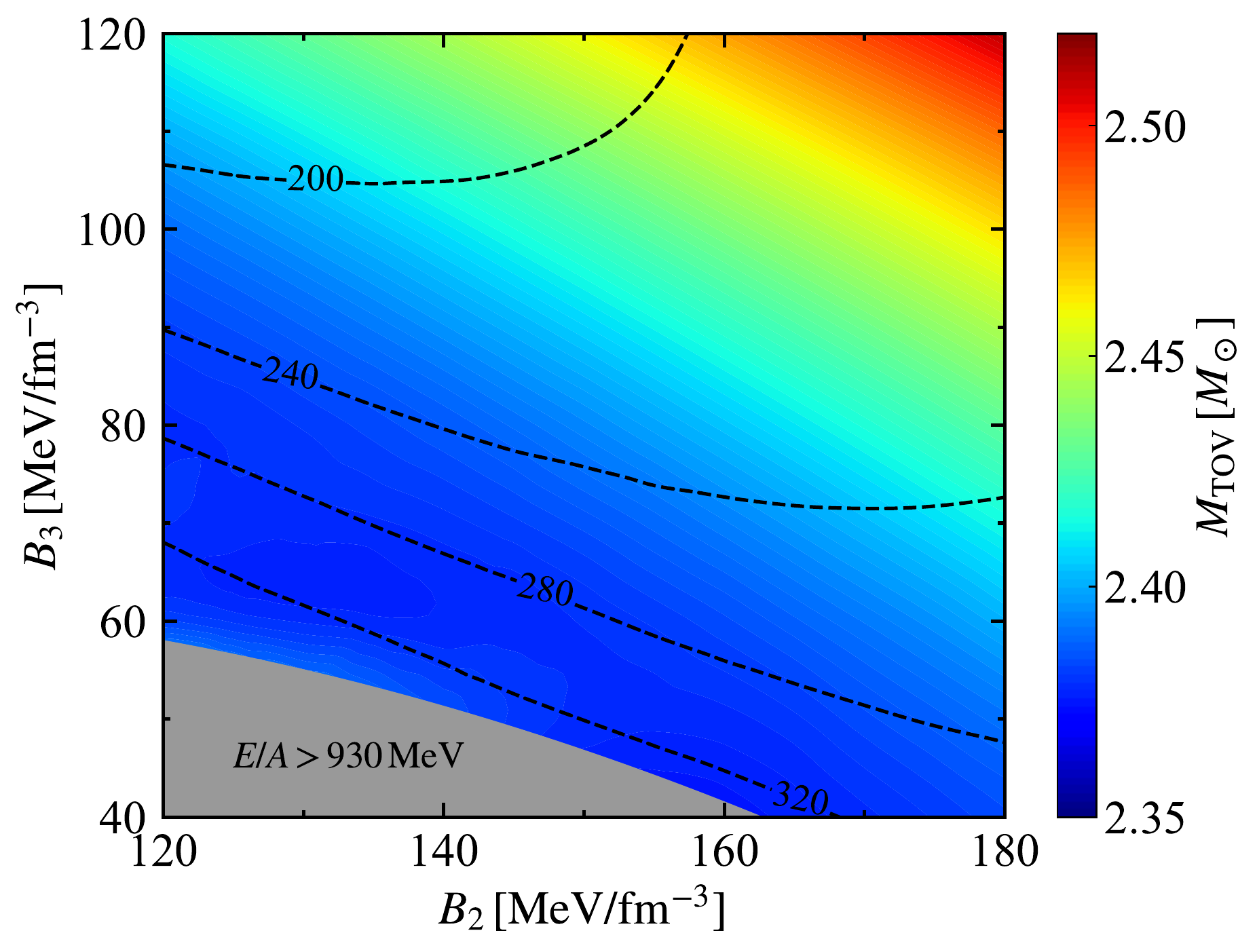}
\caption{Maximum mass $M_{\rm TOV}$ and tidal deformability of a typical $1.4\,M_\odot$ star ($\Lambda_{1.4}$) for strangeon stars with $N_{\rm q}=9$. The invariant strange quark mass is $\hat m_s = 280\,{\rm MeV}$. Contours of $M_{\rm TOV}$ are illustrated in colors, while contours of $\Lambda_{1.4}$ are plotted in black dashed lines. The lower-left grey region is ruled out since strangeon matter becomes unstable with the minimum energy per baryon $E/A > 930\,{\rm MeV}$.}
\label{fig:contour_M_L14}
\end{figure}

To investigate the parameter dependence more carefully, in Fig.~\ref{fig:contour_M_L14} we present contours of maximum mass and tidal deformability for strangeon stars. It is found that the maximum mass increases with $B_2$ and $B_3$, which even exceeds $2.5\,M_\odot$ in the top right corner. In light of the recent measured massive compact object (2.50-$2.67\,M_\odot$) in a compact binary coalescence of GW190814~\cite{LVC2020_ApJ896-L44}, the object may in fact be a strangeon star instead of a black hole. Meanwhile, even in the lower left corner, the maximum mass remains higher than $2\,M_\odot$, which fulfills the recent observational constraints of the massive stars: PSR J1614-2230 ($1.928 \pm 0.017\ M_\odot$)~\cite{Demorest2010_Nature467-1081, Fonseca2016_ApJ832-167}, PSR J0740+6620 ($2.08 \pm 0.07\ M_\odot$)~\cite{Fonseca2021_ApJ915-L12}, and PSR J0348+0432 ($2.01 \pm 0.04\ M_\odot$)~\cite{Antoniadis2013_Science340-1233232}. Since the central densities of $1.4\,M_\odot$ strangeon stars are much smaller than that of the most massive ones, the tidal deformability $\Lambda_{1.4}$ is insensitive to $B_2$ and is decreasing slightly with $B_3$. In the parameter space indicated in Fig.~\ref{fig:contour_M_L14}, $\Lambda_{1.4}$ ranges from $180$ to $340$, which fulfills the GW170817 constraint~\cite{LVC2018_PRL121-161101}. Note that the most recent constraints on neutron star radius by NICER \cite{Miller2021_ApJ918-L28} are not included in our comparison, as those analyses are based on the assumption of a normal neutron star surface instead of self-bound stars~\cite{Li2021_MNRAS506-5916}.

\begin{figure}
\centering
\includegraphics[width=0.55\linewidth]{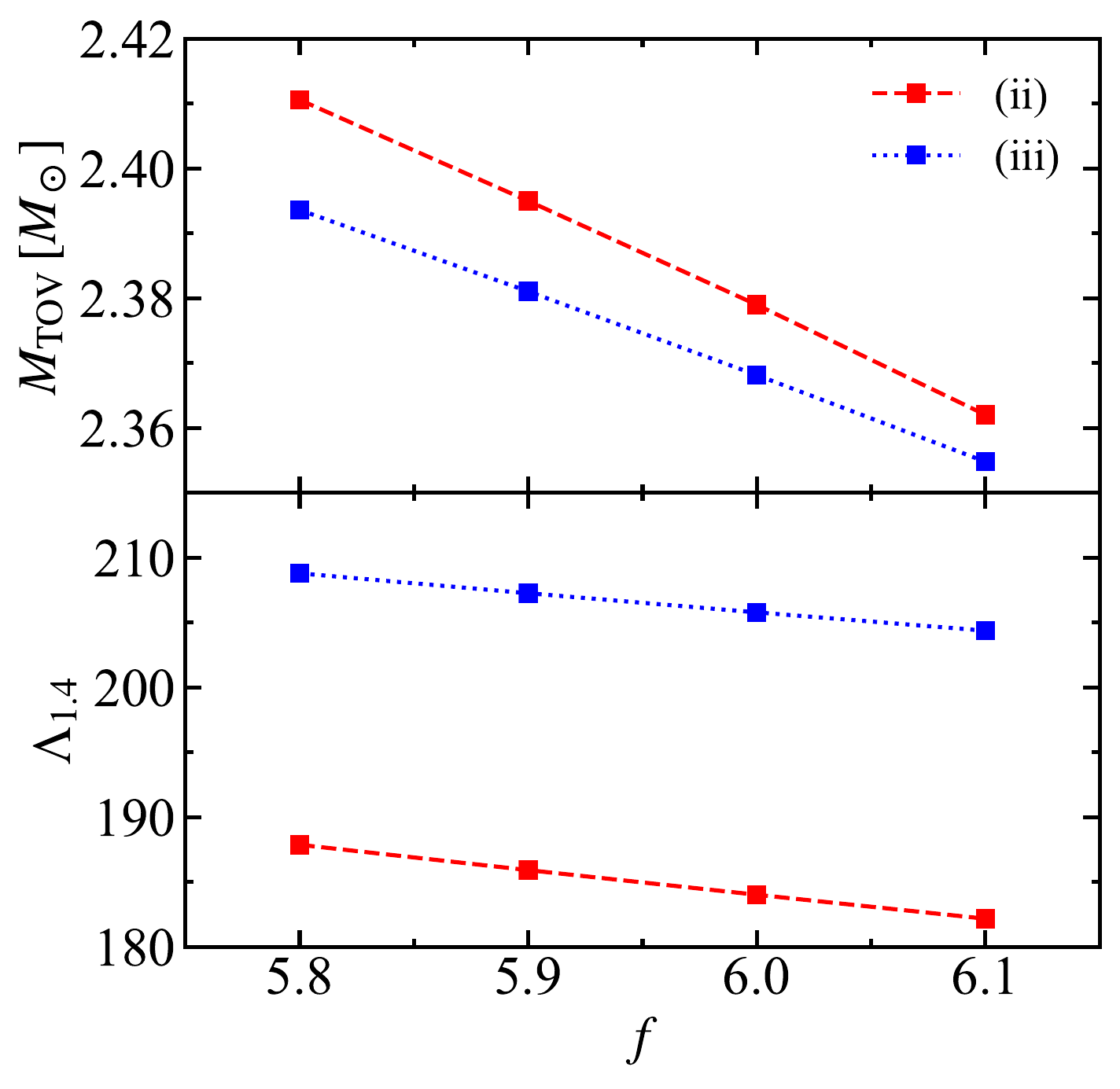}
\caption{Maximum mass $M_{\rm TOV}$ and tidal deformability ($\Lambda_{1.4}$) of strangeon stars as functions of the dampening factor $f$.}
\label{fig:Effect_of_f_on_MTOV_L14}
\end{figure}

In the discussions above, we have fixed $f=5.8$ for strangeon stars with $N_{\rm q}=9$. However, $f$ is a free parameter introduced to denote the energy level correction and the exact value of $f$ is unknown. It is thus meaningful to investigate the effects of $f$ on strangeon star structures. For this reason, in Fig.~\ref{fig:Effect_of_f_on_MTOV_L14} we present the maximum mass $M$ and tidal deformability ($\Lambda_{1.4}$) of strangeon stars as functions of the dampening factor $f$. In general, the maximum mass and tidal deformability monotonously decrease with $f$. At lower $f$, the maximum mass $M_{\rm TOV}$ may exceeds $2.4\,M_\odot$. We also notice for both cases displayed in Fig.~\ref{fig:Effect_of_f_on_MTOV_L14}, $\Lambda_{1.4}$ lies in the range of GW170817 constraint~\cite{LVC2018_PRL121-161101}. In a word, there exists a large parameter space for $f$ that the linked bag model predicts compact star structures satisfying the observational constraints on mass and tidal deformability.

\section{Discussions and Conclusions}\label{sec:conclusion}

The nature of gravity-compressed baryon matter created after core-collapsed supernova is investigated in this work, where both 2-flavoured nucleon and 3-flavoured strangeon matters are modeled with linked bags.
At this moment, pulsars are usually thought to be conventional neutron stars, while they could be strange quark stars if Witten's conjecture~\cite{Witten1984_PRD30-272} is correct. Unfortunately, it is still challenging to prove or disprove this conjecture because of the non-perturbative behavior of strong interaction. 3-flavoured quarks could be grouped in strange-clusters/strangeons if the coupling between quarks is still strong enough, where a self-bound strangeon star can be formed.
This speculative view of strangeon star has been supported by latter astronomical observations, particularly the discovery of massive radio pulsars around $2M_\odot$~\cite{Demorest2010_Nature467-1081,Fonseca2016_ApJ832-167,Antoniadis2013_Science340-1233232} since the EOS of strangeon matter is very stiff~\cite{Lai2009_MMRAS398-L31}.
A strangeon star model of pulsar glitch was proposed~\cite{Lai2018_MNRAS476-3303, Wang2020_MNRAS500-5336}, and the shear modulus of strangeon matter is constrained to be order of $10^{34}$erg/cm$^3$ in order to explain the glitch activity. In addition to the glitch phenomenon, a recent hot topic of fast radio bursts could also be interesting events to reveal the magnetospheric activity of strangeon stars~\cite{Wang2018_ApJ852-140, Luo2020_Nature586-693}. Note that if atomic line feature would be discovered in X-ray spectrum of radio pulsars, the strangeon star model has to be ruled out~\cite{Xu1999_ApJ522-L109}.
It is, therefore, urgent to model strangeon matter with microscopic foundation in consistent with nuclear physics, in order to predict effectively astronomical observations in the future.
Here we try to do so with a linked-bag model, as a first step, in the regime of non-perturbative QCD.

In this paper, we model the strong condensed matter of 3-flavoured strangeons with a linked bag approach. For fixed bag parameters $B_2$ and $ B_3$, the model parameters $C_1$ and $z_0$ are calibrated by reproducing the saturation properties ($E/A$, $K_0$, $E_{\rm sym}$ and $L$) of nucleon matter. Beside these, a dampening factor $f$ is introduced to account for the reduction of quark energy level corrections. The obtained energy per baryon of strangeon matter is usually smaller than that of nuclear and hyperon matter, which can be further reduced if we adopt larger quark numbers ($N_{\rm q}$) inside a strangeon. The corresponding EOSs of strangeon matter become stiffer as well, which increases the maximum mass of strangeon stars. It is found that, for $N_{\rm q}=9$, the maximum mass of strangeon stars could be $\sim 2.5 M_\odot$, while the tidal deformability of a $1.4\,M_\odot$ strangeon star $\Lambda_{1.4}\simeq (180$-340). To investigate the parameter dependence, the maximum mass and tidal deformability of strangeon stars predicted by the linked bag model are examined by adopting various $B_2$, $ B_3$, and $f$, which are consistent with the current astrophysical constraints in a large parameter space. More refined theoretical efforts are required in our future study, where the quark single particle energy~\cite{Zhang1992_PRC46-2294}, the interactions among quarks (instanton, electric and magnetic gluon exchange, etc.), the center-of-mass correction~\cite{Bartelski1984_PRD29-1035}, the effects of color superconductivity~\cite{Buball2005_PR407-205, Alford2008_RMP80-1455}, the quark composition of strangeons, and the possible mixing of different types of strangeons and baryons should be examined carefully. Those effects could easily alter our predictions on $M_\mathrm{TOV}$ and $\Lambda_{1.4}$ of strangeon stars, which should be tested further in the era of multi-messenger astronomy.

\section*{ACKNOWLEDGMENTS}
We would like to thank Prof. Guangshan Tian for discussion relevant to normal condensed matter, Prof. Ang Li and Prof. Makoto Oka for valuable comments and suggestions, and Mr. Yong Gao and Dr. Fei He for a preliminary calculation.
This work was supported by National SKA Program of China No.~2020SKA0120300, National Key R\&D Program of China (Grant No. 2017YFA0402602), the National Natural Science Foundation of China (Grant Nos.~11673002, U1531243, U1831104).

%\bibliographystyle{ws-ijmpe}
%\bibliography{strange_quark}

\end{document}